\documentclass[12pt]{article}
\usepackage{amsmath}
  \usepackage{graphicx,xcolor,float}
\usepackage{indentfirst}
\usepackage{multirow}
\usepackage{textcomp}
\usepackage{hyperref}
\usepackage{amssymb}
\usepackage{amsthm}
\usepackage{overpic}
\usepackage{bbm,bm,epsfig,ulem}
\usepackage{url}
\usepackage{caption}
\usepackage{subfigure}
\usepackage{cite}

\begin{document}

\begin{center}
{\Large\bf The Glauber model and flow analysis with Pb-Pb collisions at $ \sqrt{s_{\rm NN}}$=2.76 TeV}
\end{center}

\vspace{0.2cm}

\begin{center}
{\bf Ce-ran Hu$^{1 ,2}$}
\footnote{E-mail: huceran0924@berkeley.edu},
{\bf Wen-tao He$^{3}$}
\footnote {E-mail: hewentao173@163.com},
{\bf Cheng-ze Lyu$^{4}$}
\footnote{E-mail: johnson041113@163.com},
{\bf Kang-jie Yang$^{5}$}
\footnote{E-mail: 20230164@tsinglan.com},
{\bf Lily Yao$^{6}$}
\footnote{E-mail: lilyyao\_girl@outlook.com},
{\bf Rui-chong Zhao$^{7}$}
\footnote{E-mail: ruichongzhao@gmail.com}

{\small $^{1}$School of Physical Sciences,
University of Chinese Academy of Sciences, Beijing, China\\
 $^{2}$Department of Physics, University of California, Berkeley, USA\\
 $^{3}$WLSA Shanghai Academy, Shanghai, China}\\
 $^{4}$Shanghai Starriver Bilingual School, Shanghai, China\\
 $^{5}$Tsinglan School,Guangdong,China\\
 $^{6}$Miss Hall's School, Pittsfield, Massachusetts, USA\\
 $^{7}$Shenzhen College of International Education, Guangdong, China\\

\end{center}

\vspace{2cm}
\begin{abstract}

 This work presents data analysis on Pb-Pb collisions with $\sqrt{s_{\rm NN}}$=2.76 TeV in centrality $40\%-50\%$. We present introduction and Monte-Carlo simulation results of the Glauber model, which shed light on the initial geometric configuration of heavy ion collisions. Three-dimensional correlation function is plotted, and Fourier decomposition is carried out in order to obtain elliptic flow. Based on the assumption that non-flow effect is less prominent in long-range area, we separate it from the second Fourier decomposition of two-particle correlation function by making polynomial curve fitting.      
\end{abstract}

\section{Introduction}

Quantum Chromodynamics (QCD) is one of the most successful theories created to extensively describe elementary particles. To be explicit, it aims at describing strong the interaction widely existing in nuclei and hadrons, where quarks and gluons play a dominant role in deciding the physical properties of the system.

One particularly interesting physical system within the framework of QCD is quark-gluon plasma (QGP)\cite{VanHove:1986mz}. It consists of quarks and gluons with ultrahigh density and temperature that it bears much resemblance to matter existed in the very beginning of the universe. Important as it is, direct insight into such system is, nevertheless, faced with extreme difficulty. The problem is basically twofold. On one hand, extreme density and temperature of QGP can only be facilitated with enormous amount of energy and harsh conditions. On the other hand, quark confinement puts stringent limit on the lifespan of QGP, approximately in the scale of femtosecond. QGP experiences hadronization afterwards, in which quarks and gluons are bounded together and confined within hadrons. Fortunately, both of questions have been properly settled nowadays. The production of QGP has been made possible with the help of heavy ion collisions on the Large Hadron Collider (LHC). As for the latter one, though direct study has not yet come into reality, we are indeed capable of analyzing data collected after hadronization to speculate or verify the physical properties of the system.

The evolutionary patterns of QGP has been discovered that, to high extent of accuracy, match the hydrodynamics of perfect liquid, with which the shear viscosity to entropy density is governed by the inequality: $\eta / s \leq \frac{1}{4\pi}$, as is suggest by anti-de Sitter/conformal field theory(AdS/CFT) conjecture \cite{Kovtun:2004de}. Besides, QGP tends to maintain its initial anisotropy, which can be classified as the effect of elliptic flow, triangular flow, etc. Such effects can be extracted from the Fourier decomposition coefficients of the correlation function calculated from the data collected by detectors, in our case, particularly, data from Compact Muon Solenoid (CMS) of Pb-Pb collisions with $\sqrt{s_{\rm NN}}$=2.76 TeV in centrality $40\%-50\%$.

The remaining parts of this paper are organized as follows: Section 2 discusses about the Glauber model, with which we simulate the initial geometric figures and distributions in heavy ion collisions. In Section 3, we present our plot of the 3-dimensional correlation function against $\Delta \phi$ and $\Delta \eta$. Several features of the figure are noted and analyzed. Moreover, the Fourier decomposition is carried out and analysis on the results is presented. In Section 4, we manage to separate non-flow effect from the physical flow effect in the second coefficient of Fourier decomposition, $\rm v_2^2$. Section 5 aims at providing an overall summary of the text and proposes possible future research direction.  

\section{The Glauber model}
It should be clarified that with current experimental apparatus and methods, we are unfortunately not capable of directly probing the primary geometric states of the collision events. However, technological backwardness has in turn stimulated the development of several simulation methods with the help of computers. The Glauber model, has been tested to be one of the most successful methods that shed light on our physical intuition on the initial geometry. It is based on the approximation that collisions are independent, sequential, and is built upon scattering theories in Quantum Field Theory. It aims at tackling high energy scatterings with composite particles, such as nucleon-nucleon scatterings on LHC. In this work we present the "Glauber Monte Carlo" Model(GMC), in which nucleus is modeled by a group of uncorrelated nucleons based on a given density distribution. In this model"Optical limit"is utilized to make the integration numerically feasible\cite{Miller:2007ri}.

The Glauber model is capable of simulating several physical properties for the initial state, including nuclear charge density $\rho$, number of participating nucleons, $N_{part}$, number of colliding nucleons, $N_{coll}$, and the distribution of colliding nucleons. The sole parameter required to be introduced is a randomly generated impact parameter b.

When it comes to nuclear charge density $\rho$, several empirical distribution functions have been constructed to adapt to different nuclei. Three types of density distribution are pointed out here\cite{Alver:2008aq}:
\begin{eqnarray}
 \rho(r)=\rho_0\frac{1+\omega(r/R)^2}{1+e^{\frac{r-R}{a}}}, \\
 \rho(r)=\rho_0\frac{1+\omega(r/R)^2}{1+e^{\frac{r^2-R^2}{a^2}}},\\
 \rho(r)=4\hspace{0.05cm} \rho_0 \hspace{0.05cm}e^{-\alpha r}({\frac{\cosh{(-\beta r)}}{r}})^2
\end{eqnarray}

 Eq.(1) is a Fermi-like distributions with four parameters. $\rho_0$ stands for the nucleon density in initial state, $\omega$ is a measurement of deviation from a perfect sphere. R is the radius of the nucleus, while a is a characteristic length. The same parameterization works for Eq.(2), the form of a Gaussian function. As for Eq.(3), the Hulthen formula, $\alpha=2.807$ $\rm fm^{-1}$ and $\beta = 0.9465$ $\rm fm^{-1}$ are two lengths calculated from experimental results. Of all empirical equations, Eq.(1) can be appropriately applied to most nuclei used in collision experiments. Eq.(2) is especially designed for sulfur while Eq.(3) is parameterized for deuteron\cite{Alver:2008aq}.

The rest of this section concentrates on our simulated results in the Glauber model. 

\begin{figure}[htbp]
\centering 
{
\includegraphics[scale=0.4]{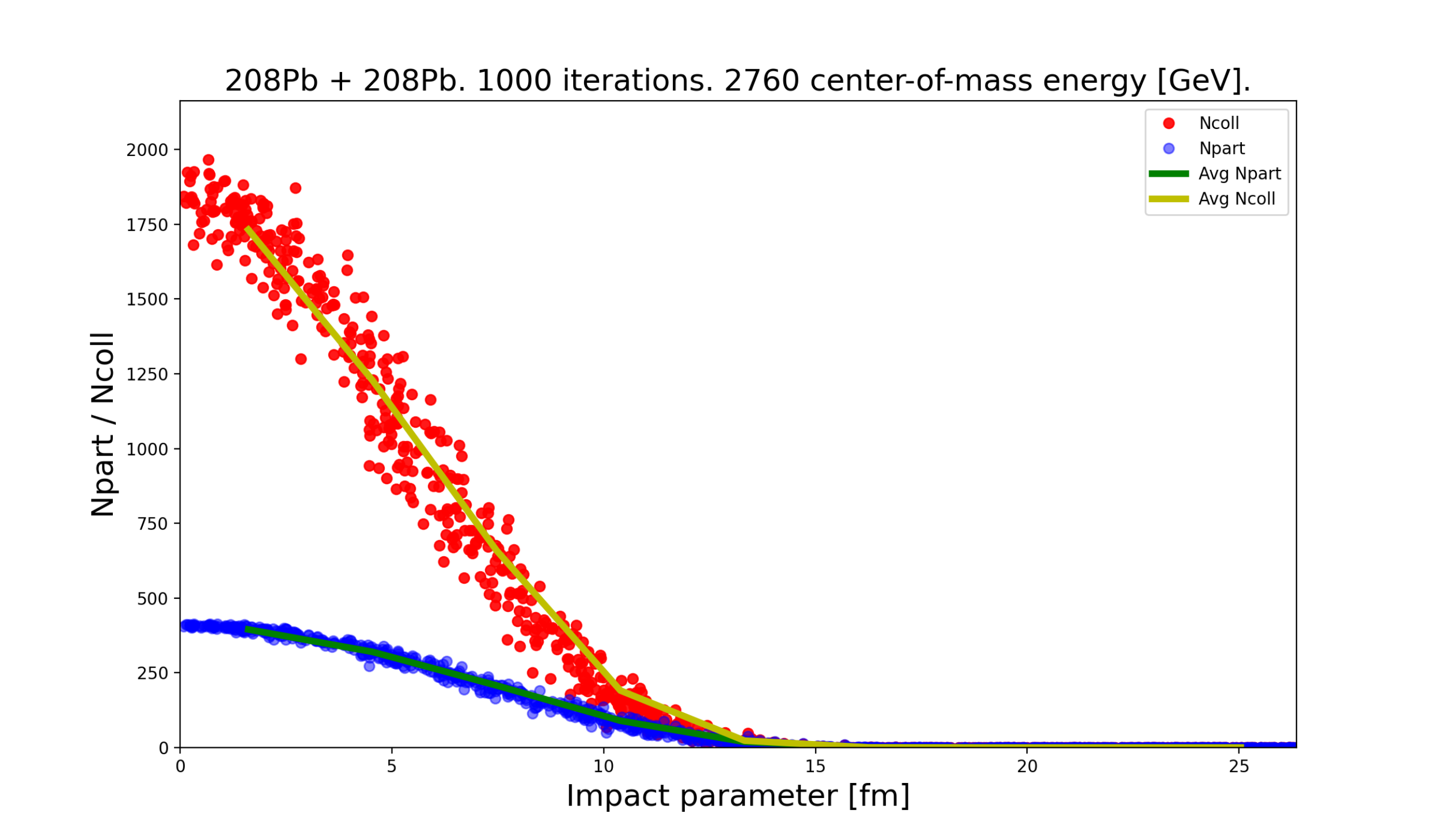}
}

\caption{$N_{coll}$ and $N_{part}$ versus impact parameter b with Pb-Pb collision at $\sqrt{s_{\rm NN}}=$2.76 TeV. The scatter points are fitted to the average curves in green and yellow.}
\label{NNNN}
\end{figure}

On Figure~\ref{NNNN}, a display of the values of $N_{coll}$ and $N_{part}$ are shown. Both values show a reasonable descend to zero as impact parameter b increases.

\begin{figure}[htbp]
\centering 
\begin{minipage}[t]{0.48\textwidth}
\centering
\includegraphics[scale=0.4]{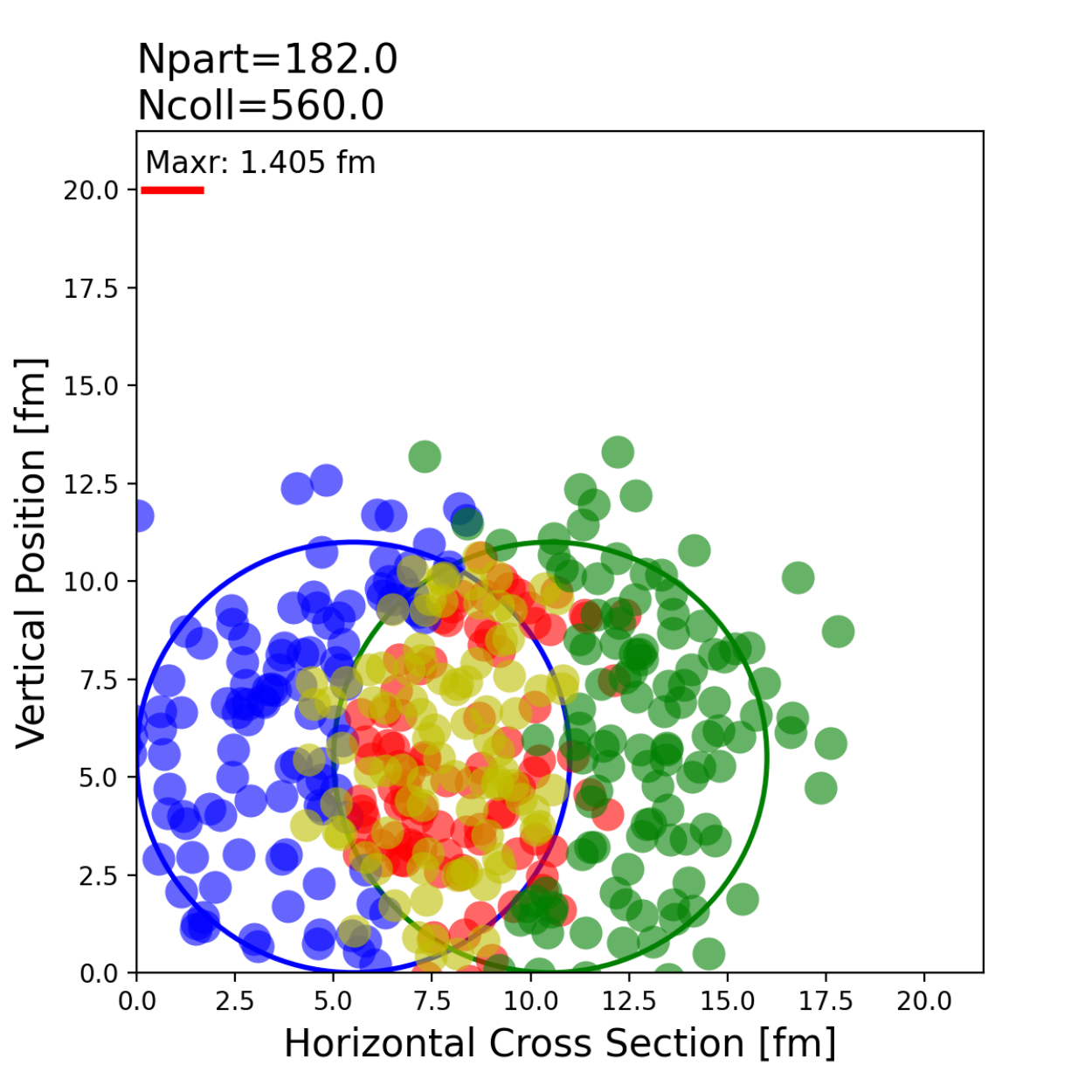}
\caption{Beam-in View}
\label{2}
\end{minipage}
\begin{minipage}[t]{0.48\textwidth}
\centering
\includegraphics[scale=0.4]{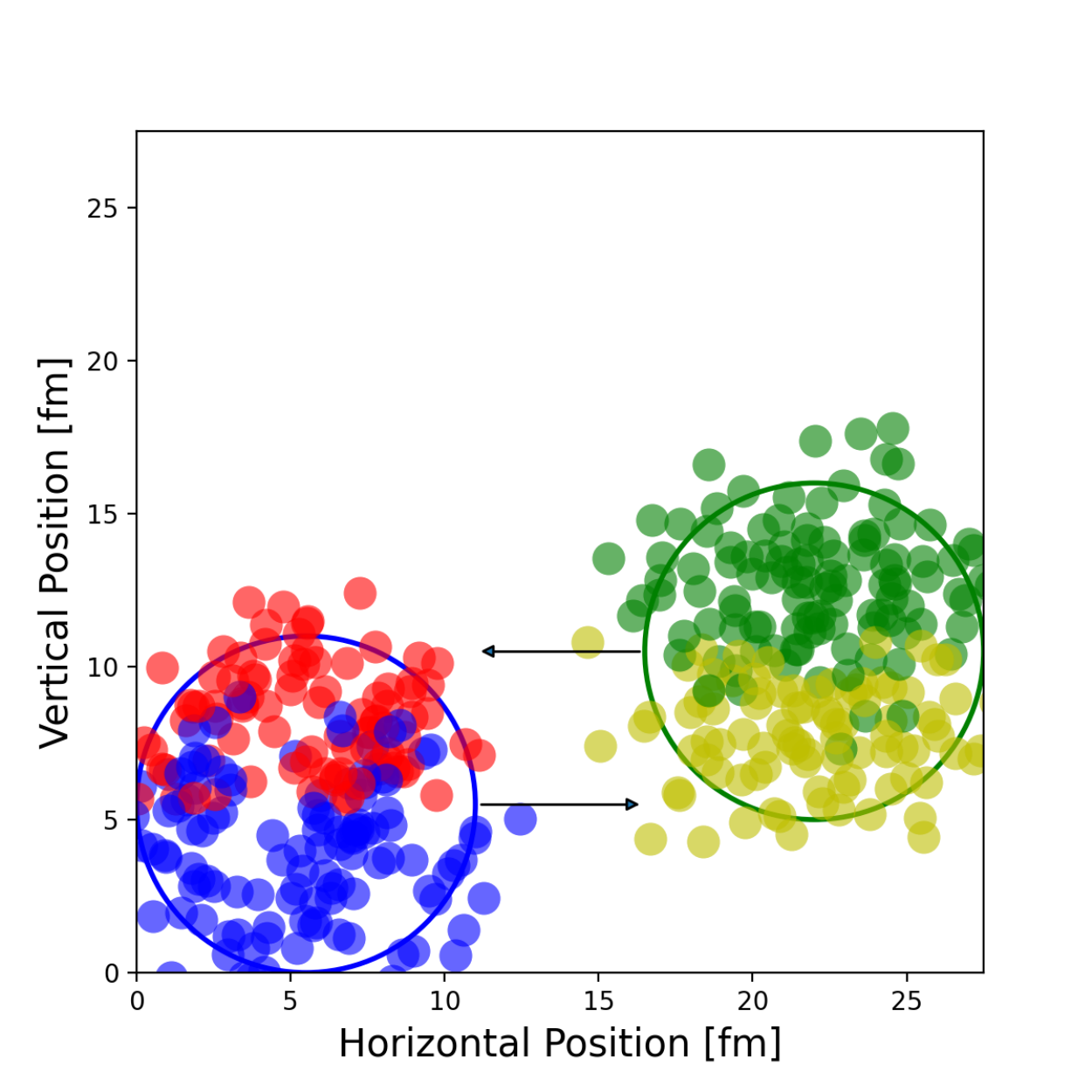}
\caption{Side View}
\label{3}
\end{minipage}
\end{figure}

After nuclear density $\rho$,  $N_{coll}$ and $N_{part}$ being set up, the of rest of work is to input the all these parameters to generate the initial geometry in a particular event. A typical example is presented in Figure~\ref{2} and Figure~\ref{3}, in which we try to present collisions from two rather different  views, beam-in view and side view. The green and blue symbols serve as spectators, while red and yellow ones participate in nucleon-nucleon collisions.

One specific parameter matters when it comes to whether the collision could occur or not. The "ball diameter" is defined as\cite{dEnterria:2020dwq} :
\begin{eqnarray}
D=\sqrt{\frac{\sigma_{NN}}{\pi}}
\end{eqnarray}

$\sigma_{NN}$ refers to the inelastic nucleon-nucleon cross section. Uncertainty arises from the unknown elastic cross section. $\sigma_{NN}$  solely depends on the collision energy. In this model, it is assumed that the chance for a  binary collisions between any two nucleons with their relative transverse distance $d\textgreater D$ is ruled out.

\section{Correlation and elliptic flow}
\begin{figure}[htbp]
\centering
{
\includegraphics[scale=1]{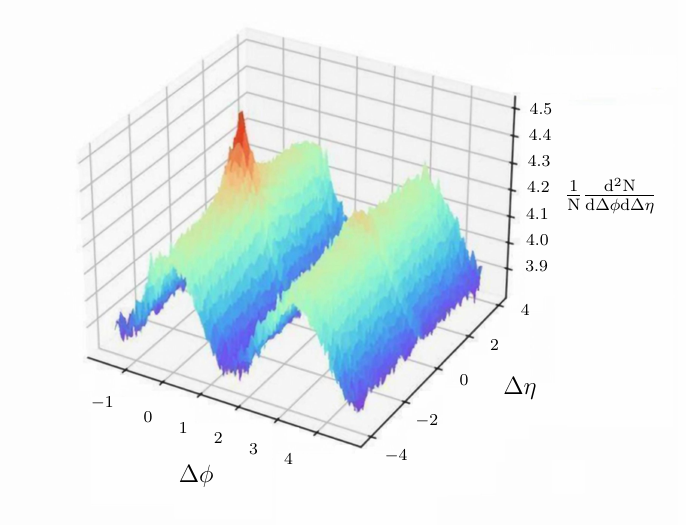}
}
\vspace{-0.8cm}
\caption{The three-dimensional plot of correlation function against $\Delta \phi$ and $\Delta \eta$. $\Delta \phi$ ranges from $-\frac{\pi}{2}$ to $\frac{3\pi}{2}$ in order to match previous work. $\Delta \eta$ varies within -4 to 4, approximately.}
\label{Cor}
\end{figure}

One piece of  interesting information we can extract from the data is the correlation function against differences with azimuthal angle  $\Delta \phi$, and differences with pseudorapidity $\Delta \eta$. Here peudorapidity $\eta$ is defined as $\eta=-\ln[\tan(\frac{\theta}{2})]$, where $\theta$ is the polar angle. The correlation function aims at providing an insight into the interaction between particles produced in the same collision event from a statistical point of view. Namely, the correlation function is obtained by dividing signal by background $\frac{S(\Delta \phi,\Delta \eta)}{B(\Delta \phi,\Delta \eta)}$. Signal is defined by making histograms about the $\Delta \phi$ and $\Delta \eta$ from traces within one event and accumulate for all events, while background is plotted by making histograms with $\Delta \phi$ and $\Delta \eta$ from traces of different events. Background is normalized so that it matches the scale of signal. By definition, correlation function is proportional to the true-pair distribution\cite{PHENIX:2008osq}.  It is straightforward to test that correlation function is equivalent to the expression $\rm \frac{1}{N}\frac{d^2 N}{d\Delta \phi d\Delta \eta}$. Its plot is shown in Figure~\ref{Cor}.

Several features of  Figure~\ref{Cor} are worth of mentioning: (1)With centrality $40\%-50\%$, we could discover a symmetry of the configuration with respect to line $\Delta \eta=0$. (2)The configuration remains approximately stable along the direction of $\Delta \phi$ axis (except for the point $(\Delta \phi,\Delta \eta)$=(0,0)). (3)A spike-like structure locates around the coordinate $(\Delta \phi,\Delta \eta)$=(0,0), indicating that  outcoming particles tend to stay close. Such phenomena can be properly reasoned by Bose-Einstein Correlations. Bose-Einstein Correlations come from the fact that elementary particles are associated with waves, where interference takes place to create correlation. For bosons, they tend to form a compact bunch under such correlation\cite{Glauber:1963tx}. (4)There are two ridges in the figure. One is along $\Delta \phi=0$. It implies the existence of correlations between particles from  single jets. Another ridge, on the other hand, locates far from the center along with $\Delta \phi=\pi$, which corresponds to correlations of back-to-back jets.

As is mentioned in Section 1, QGP has the tendency to maintain its initial anisotropy, which can be unfolded by studying the Fourier decomposition coefficient of the azimuthal angle correlation function, $\rm \frac{dN}{d\Delta\phi}$. 
\begin{eqnarray}
\rm \frac{dN}{d\Delta\phi}=v_0+\sum_{n=1}^{\infty}v_n\cos{[n\Delta \phi]}
\end{eqnarray}

\begin{figure}[htbp]
\centering
{
\includegraphics[scale=0.9]{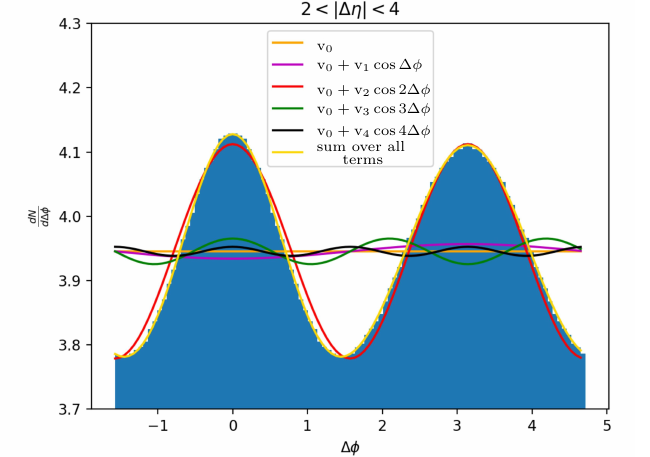}
}
\vspace{-0.3cm}
\caption{The azimuthal angle correlation function and the components of different orders of Fourier decomposition.}
\label{Fourier}
\end{figure}

 Here specifically, we study the second Fourier decomposition coefficient, $\rm v_2$. It is dominated by the effect of the elliptic flow, which mostly makes up the correlation function $\rm \frac{dN}{d\Delta\phi}$ \cite{Alver:2010gr}. However, due to the presence of non-flow effect contributed by secondary decays and emissions, it is inappropriate to make a histogram against $\Delta \phi$ and then decompose directly. Fortunately, non-flow effect is most prominent within short-range area, we instead select results with $2 \textless \vert \Delta \eta \vert \textless 4$ to plot histogram against $\Delta \phi$ in order to exclude the interference of non-flow effect. Detailed analysis on separating non-flow effect from elliptic flow will be made in Section 4. The outcome is presented in Figure~\ref{Fourier}. 

As is shown in Figure~\ref{Fourier}, the oscillation pattern of $\rm v_0+v_2\cos{2\Delta \phi}$ term is highly consistent with that of the correlation function. The rest of Fourier component terms, with minor amplitudes, serve as the modifications for $\rm v_0+v_2\cos{2\Delta \phi}$ term to the correlation function. Such hierarchy of Fourier components implies that elliptic flow predominates over other flows in the data set, including directed flow, triangular flow and rectangular flow, which which can be extracted from terms of $\rm v_0+v_1\cos{\Delta \phi}$, $\rm v_0+v_3\cos{3\Delta \phi}$, $\rm v_0+v_4\cos{4\Delta \phi}$, respectively.

\section{Separation of non-flow}

This section is intended for estimating and separating the non-flow contributions from physical flow with two-particle correlation functions. This time, however, instead of presenting the correlation function in the $(\Delta \eta, \Delta \phi)$ parameter space, we present $v^2_2$ in $(\eta_1,\eta_2)$ parameter space to approach the problem from another point of view.

The Fourier decomposition coefficient, $v^2_2$ of correlation function $ C(\eta_1,\eta_2,\Delta \phi)$ in such parameter space is defined as follows:\cite{PHOBOS:2010ekr}
\begin{eqnarray}
    C(\eta_1,\eta_2,\Delta \phi)=\langle \frac{\rho^{foreground}(\eta_1,\eta_2,\Delta \phi)}{\rho^{background}(\eta_1,\eta_2,\Delta \phi)}-1\rangle\\
    v^2_2(\eta_1,\eta_2)=\frac{1}{\pi}\int^{2\pi}_{0} C(\eta_1,\eta_2,\Delta \phi) \cos2\Delta \phi \hspace{0.1cm}d\Delta \phi
\end{eqnarray}

$\rho^{foreground}(\eta_1,\eta_2,\Delta \phi)$ stands for foreground pair distribution, which is calculated by making histograms of pair of traces within same events. $\rho^{background}(\eta_1,\eta_2,\Delta \phi)$, on the other hand, represents the background pair distribution, which is obtained by histogramming pair of traces from different events. The triangular brackets indicate averages over all events.

It is noteworthy to mention that in this case, correlation function in the ($\eta_1$,$\eta_2$) space does not take the effect of detector acceptance, secondary decay and scattering into consideration. Further discussions about the influences from such factors on correlation function are included in\cite{PHOBOS:2007whv}.

\begin{figure}[htbp]
\centering
{
\includegraphics[scale=0.9]{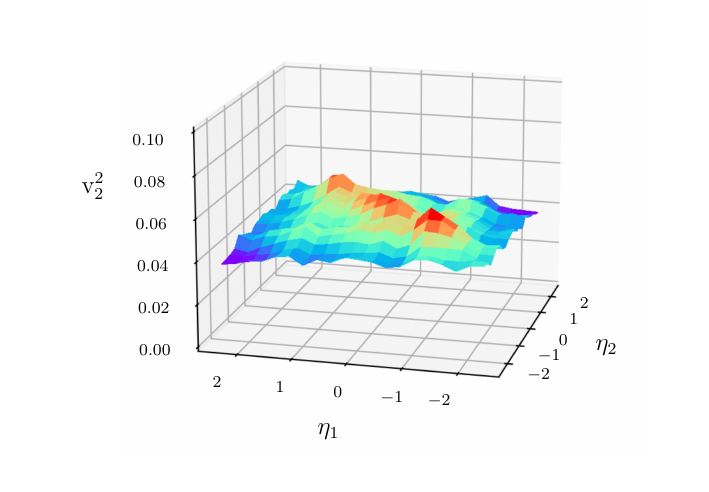}
}
\vspace{-0.3cm}
\caption{ The two-particle correlation function in parameter space $(\eta_1,\eta_2)$ with centrality $40\%-50\%$ in Pb-Pb collision at $\sqrt{s_{\rm NN}}$=2.76 TeV.
.}
\label{total}
\end{figure}

The plot of the correlation function is plotted in Figure~\ref{total}. In detail, we collect $\eta_1$ and $\eta_2$ ranging from -2.5 to 2.5. We divide the parameter space $(\eta_1,\eta_2)$ into 20 $\times$ 20 grids. In each grid we perform the Fourier decomposition of correlation function. As is clearly presented in Figure~\ref{total}, a ridge shows up along the line $\eta_1=\eta_2$. It corresponds to the Bose-Einstein Correlations mentioned in Section 3, or the spike-like configuration in  Figure~\ref{Cor}. Meanwhile, $v^2_2$ is less structured in the rest part of the parameter space, approximately maintaining at a constant value. Such differences between two regions can be reasonably attributed to the fact that non-flow effect is most remarkable in short-range area.

The rest of this section mainly focuses on the idea and results of separating non-flow from physical flow.

Without loss of generality, in order to separate non-flow and physical flow, one has to make an assumption that non-flow and physical flow do not couple with each other, otherwise the situation can be much more complicated. As a result, we arrive at:
\begin{eqnarray}
v^2_2(\eta_1,\eta_2)=v^{*\hspace{0.1cm}2}_2(\eta_1,\eta_2)+\epsilon(\eta_1,\eta_2)
\end{eqnarray}

In Eq.(8), $v^{2\hspace{0.1cm}*}_2$ refers to the contribution from physical flow to $v^2_2$, while $\epsilon$  represents the contribution of the non-flow term.\cite{Poskanzer:1998yz}

Furthermore, as for $v^{*\hspace{0.1cm}2}_2$, it can be divided into $v^*_2(\eta_1) \hspace{0.1cm}\times v^*_2(\eta_2)$, given the fact that the correlation created by physical flow solely depends on pseudorapidity $\eta$. 

Based on these assumptions, the rest of the work is left to make a suitable fit for the physical flow in a specific area to accomplish the separation. Keep in mind that non-flow effect is less significant in long-range area.  As a consequence, we make a fit for the physical flow in $|\eta_1-\eta_2|\textgreater 2$ in the parameter space and expand the fitting result to the whole parameter space we investigate. In our case, polynomial of fourth degree has been chosen to approximate $v^*_2(\eta)$. Indeed, we have to confess that it is a rough fit for physical flow due to the lack of data as well as our fitting method. It does, nonetheless, provide a primary insight into the role physical flow plays. 

\begin{figure}[htbp]
\centering
{
\includegraphics[scale=0.9]{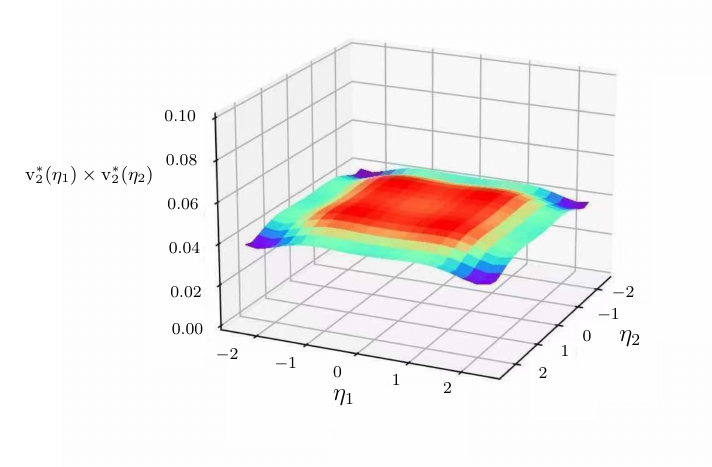}
}
\vspace{-0.3cm}
\caption{ Physical flow's contribution to $v^2_2$ fitted by degree 4 polynomial in region $|\eta_1-\eta_2|\textgreater 2$ and expanded to the whole parameter space.}
\label{physical}
\end{figure}

As is presented in Figure~\ref{physical}, the configuration of physical flow term is rather plain, except for four corners in the figure, where slopes are identified because of the polynomial fitting method. Such flat configuration matches our expectation that the effect of physical flow is distributed almost evenly in the parameter plane.

\begin{figure}[htbp]
\centering
{
\includegraphics[scale=1.8]{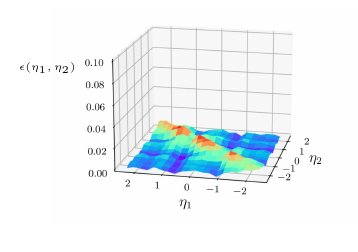}
}
\vspace{-0.8cm}
\caption{ Non-flow correlation $\epsilon(\eta_1, \eta_2)$ obtained by subtracting the two-particle correlation function $v^2_2(\eta_1,\eta_2)$ by the physical flow's contribution to  $v^*_2(\eta_1) \hspace{0.1cm}\times v^*_2(\eta_2)$.}
\label{non}
\end{figure}

With Eq.(8), we are allow to obtain the non-flow effect on $v^2_2$ by subtracting $v^2_2(\eta_1,\eta_2)$ in Figure~\ref{total} by the approximated physical term $v^*_2(\eta_1) \hspace{0.1cm}\times v^*_2(\eta_2)$ in Figure~\ref{physical}. The configuration of $\epsilon(\eta_1,\eta_2)$ is demonstrated in Figure~\ref{non}. A few comments are in order. Initially, a ridge-like region lies along $\eta_1=\eta_2$, indicating the strong non-flow correlation corresponding to the Bose-Einstein Correlation as is discussed earlier. Besides, in the rest of the parameter space, the non-flow correlation fluctuates around 0, which perfectly illustrates that its effect is nearly negligible in long-range area.

In this paper, efforts are made to separate non-fow from physical flow. However, eliminating the effect of non-flow does not complete the error analysis. Here we provide a phenomenological formula for error analysis:
\begin{eqnarray}
\sigma^{2}_{v_2^{obs}}=(f\times\sigma_{\epsilon_2})^2+(\sigma_{stat})^2+(\sigma_{non})^2
\end{eqnarray}

Eq.(9) is constructed under the assumption that error effects are independent from each other. $\sigma_{\epsilon_2}$ is the fluctuation of initial geometry, f is a factor connecting the initial and final states. $\sigma_{stat}$ is the statistical fluctuation, and $\sigma_{non}$ refers to non-flow effect. As is mentioned, the separation of the last term has been presented in this paper. Hopefully in the future studies on separation of
$(f\times\sigma_{\epsilon_2})$ and $\sigma_{stat}$ can be done so that it will increase the capability of extracting physical information from experimental data.

\section{Conclusions}
We have presented a brief introduction to the Glauber model and simulated results generated by computer simulations. With open data of Pb-Pb heavy ion collisions at $ \sqrt{s_{\rm NN}}$=2.76 TeV from CMS detector with centrality $40\%-50\%$, correlation function $\rm \frac{1}{N}\frac{d^2 N}{d\Delta \phi d\Delta \eta}$ in $(\Delta \eta,\Delta \phi) $ space is presented, and the Fourier decomposition with traces $2 \textless \vert \Delta \eta \vert \textless 4$ is calculated. Important facts include, Bose-Einstein Correlation is obvious on $(\Delta \eta,\Delta \phi)$ space, which corresponds to the point $(0,0)$. Besides, single jets and back-to-back jets also generate significant effect. Two-particle correlation $C(\eta_1,\eta_2,\Delta \phi)$ in $(\eta_1,\eta_2)$ plane is defined, and the Fourier decomposition coefficient $v^2_2$ is extracted, with which we perform the separation of non-flow from physical flow. As for the two-particle correlation on the $(\eta_1,\eta_2)$ plane, we have made a polynomial fit for the physical flow in long-range area in order to separate it from non-flow. The take-home massage is, physical flow is highly plain and flat, with only slopes at the corners because of the polynomial fitting method. Non-flow is most severe in short-range area, or namely, along the line $\eta_1=\eta_2$, and oscillating around zero in other areas.

In conclusion, the main focuses of this paper are simulation of initial geometry and interpreting data collected after hadronization. Both of them matter in search for understanding QGP’s the characteristic of maintaining initial anisotropy, thus understanding the overall physical properties of QGP.

\section*{Acknowledgements}
We would like to thank professor Gunther Roland for delivering lectures on quark-gluon plasma and enlightening us on relevant questions. We also thank Jie-ming Lin, for holding inspiring discussions and sharing interesting ideas with us.

\end{document}